\documentclass[twocolumn,superscriptaddress,showpacs,prb]{revtex4}

\usepackage{graphicx}%
\usepackage{dcolumn}
\usepackage{amsmath}
\usepackage{bm}

\begin{document}



\preprint{submitted to Physical Review Letter}
\title{
Electron correlation in FeSe superconductor studied by bulk-sensitive 
photoemission spectroscopy
}

\author{A. Yamasaki}
\author{Y. Matsui}
\affiliation{Faculty of  Science and Engineering, Konan University, Kobe 658-8501, Japan }

\author{S. Imada}
\affiliation{College of Science and Engineering, Ritsumeikan University, Kusatsu 525-8577, Japan}

\author{K.~Takase}
\author{H. Azuma}
\affiliation{College of Science and Technology, Nihon University, Chiyoda, Tokyo 101-8308, Japan}

\author{T. Muro}
\author{Y.~Kato}
\altaffiliation{Present address : Diamond Research Laboratory, AIST, Tsukuba 305-8568, Japan}
\affiliation{Japan Synchrotron Research Institute, Sayo, Hyogo 679-5198, Japan}

\author{A.~Higashiya}
\altaffiliation{Present address : Industrial Technology Center of Wakayama Prefecture, Wakayama 649-6261, Japan}
\affiliation{RIKEN, Mikazuki, Sayo, Hyogo 679-5148, Japan}

\author{A.~Sekiyama}
\author{S.~Suga}
\affiliation{Graduate School of Engineering Science, Osaka University, Toyonaka, Osaka 560-8531, Japan}

\author{M. Yabashi}
\affiliation{Japan Synchrotron Research Institute, Sayo, Hyogo 679-5198, Japan}

\author{K.~Tamasaku}
\author{T. Ishikawa}
\affiliation{RIKEN, Mikazuki, Sayo, Hyogo 679-5148, Japan}

\author{K. Terashima}
\affiliation{College of Science and Engineering, Ritsumeikan University, Kusatsu 525-8577, Japan}

\author{H.~Kobori}
\author{A.~Sugimura}
\affiliation{Faculty of  Science and Engineering, Konan University, Kobe 658-8501, Japan }

\author{N. Umeyama}
\affiliation{Department of Applied Physics, Tokyo University of Science, Shinjuku, Tokyo 162-8601, Japan}
\affiliation{Nanoelectronics Research Institute, AIST, Tsukuba 305-8568, Japan}

\author{H. Sato}
\affiliation{Faculty of  Science and Engineering, Chuo University, Bunkyo, Tokyo 112-8551, Japan}

\author{Y. Hara}
\affiliation{Ibaraki National College of Technology, Hitachinaka 312-8508, Japan}

\author{N. Miyakawa}
\affiliation{Department of Applied Physics, Tokyo University of Science, Shinjuku, Tokyo 162-8601, Japan}

\author{S. I. Ikeda}
\affiliation{Nanoelectronics Research Institute, AIST, Tsukuba 305-8568, Japan}

\date{\today}

\begin{abstract}
We have investigated the electronic structures of recently discovered superconductor FeSe by soft-x-ray and hard-x-ray  photoemission spectroscopy with high bulk sensitivity.
The large Fe $3d$ spectral weight  is located in the vicinity of the Fermi level ($E_F$), which is demonstrated to be a coherent quasi-particle peak.
Compared with the results of the band structure calculation with local-density approximation, Fe $3d$ band narrowing and the energy shift of the band toward $E_F$ are found, suggesting  an importance of the electron correlation effect in FeSe. 
The self energy correction provides the larger mass enhancement value  ($Z^{-1}\simeq$3.6) than in Fe-As superconductors and enables us to separate a incoherent part  from the spectrum. 
These features are quite consistent with the results of recent dynamical mean-field calculations, 
in which the incoherent part is attributed to the lower Hubbard band.

\end{abstract}


\maketitle

%
%

\section{Introduction}

Fe-based high-$T_c$ superconductors have attracted enormous attention for their possibly new-type superconducting mechanism and the potential of breaking the deadlock in the high-$T_c$ superconductor research field.  
A fluorine doped LaFeAsO has been discovered to be a superconductor below $T_c$= 26~K, which contains the two dimensional Fe plane in the Fe$_2$As$_2$ layer.~\cite{Kamihara_08}
So far, several tens of superconductors in limited types of mother materials such as LaFeAsO, BaFe$_2$As$_2$, and LiFeAs,
have been synthesized.~\cite{Norman_08}
In addition, a simple Fe-Se binary compound (FeSe$_{0.82}$) has been discovered to show superconductivity.~\cite{Hsu_08}
The appearance of superconductivity in the FeSe system indicates Fe$_2X_2$ ($X$=P, As, and Se) layer is essential for the superconductivity in these Fe-based  superconductors.
The density functional study has pointed out  that FeSe is not a conventional electron-phonon superconductor, being similar to LaFeAsO$_{1-x}$F$_x$ system.~\cite{Subedi_08}
Other common features to the Fe-based superconductors have also been revealed, which both the antiferromagnetic spin fluctuation and the anion height  are closely related to the appearance of the superconductivity.~\cite{Imai09,Okabe10}

There are, however, many differences between FeSe and other Fe-based superconductors: in FeSe 
(i) there is no separating layer and the Fe$_2$Se$_2$ layer is electrically neutral,
(ii) the superconductivity is very sensitive to the deviation from the stoichiometric composition,~\cite{McQueen_09,Pomjakushina09}
(iii) there is no magnetically ordered state in $p$-$T$ phase diagram.~\cite{Medvedev09}
Furthermore, it is pointed out by theoretical studies  that the electron correlation effect in FeSe is stronger than other Fe-based superconductors.~\cite{Miyake10,Aichhorn10}
In FeSe, only a few experimental results, for instance,  Sommerfeld coefficient $\gamma$ (= 5.4-9.1 mJ/mol K$^2$) have been so far reported~\cite{McQueen_09, Hsu_08} although the electron correlation effect  in other Fe-based superconductors was investigated by spectroscopic experiments in detail.~\cite{Yi09,Qazilbash09,Lu08,Yang09}.
The reason why  no angle-resolved photoemission spectroscopy experiment has been reported to quantitatively evaluate the electron correlation effect in FeSe is that a high quality single crystal is difficult to be grown.
Recently, Fe(Se,Te) system has been intensively investigated due to the success of the high-quality single crystal growth.
Even so, the Fe(Se, Te) system is essentially different from the end member FeSe in the sense that the magnetically  ordered state and remarkably large $\gamma$ value (= 39 mJ/mol K$^2$) have been found.~\cite{Khasanov09,Sales09}

We have examined two different synthesis processes for the FeSe superconductor and  performed the soft-x-ray and hard-x-ray photoemission spectroscopy (SXPES and HAXPES)  in order to quantitatively evaluate the electron correlation effect in the bulk.
The SXPES and HAXPES have been widely recognized as  the powerful techniques which can reveal bulk electronic structures due to the long inelastic mean free path of photoelectrons excited by high-energy x ray.~\cite{Wescke_91,Sekiyama_nature, Mo_03,yamasaki_0407, sekiyama_SP}
It is found in the angle-integrated PES spectrum that a large Fe $3d$ spectral weight is located in the vicinity of the Fermi level ($E_F$) and it decreases steeply toward $E_F$, being a similar feature to those in the other Fe-based superconductors non-doped LaFePO and LaFe{\it Pn}O$_{0.94}$F$_{0.06}$ ({\it Pn}=P, As).~\cite{Lu08,Malaeb_08}
Considering the self energy correction to the results of band structure calculations, the experimentally observed 
band narrowing and the energy shift of the band toward $E_F$ are fully explained. 
The correction also provides the renormalization factor $Z$ of $\simeq$0.28 and enables us to separate  the incoherent part of the quasi-particle spectrum.

\section{experimental}

For SXPES and HAXPES measurements, single-crystalline tetragonal [Tet(Single)] FeSe  and polycrystalline tetragonal [Tet(Poly)] FeSe were employed, together with polycrystalline hexagonal [Hex(Poly)] FeSe (likely, Fe$_7$Se$_8$) as a reference material.\cite{McQueen_09}
The Tet(Single) FeSe was grown by the chemical vapor transport method using Fe and Se powders.~\cite{Hara_09}
It was found to contain the hexagonal phase around the crystal edge by x-ray powder diffraction (XRD)  measurements.
The coexistence of two stable phases has also been reported by other groups.~\cite{McQueen_09, Mizuguchi_08,Zhang_09,Bendele10}
The spot size of the SX beam in PES measurements is small ($\sim10\mu{\rm m}\times 100\mu{\rm m}$) enough to exclude signals from the hexagonal phase near the crystal edge because of the much larger sample size ($\sim500\mu{\rm m}\times 500\mu{\rm m}$).
The Tet(Poly) FeSe  was synthesized using a high-frequency induction furnace. The furnace is very useful for quick cooling of the samples because this furnace can heat only the samples. This makes the pass time through other low temperature phase relatively short. The Hex(Poly) FeSe  was prepared by a conventional solid state reaction.
Crystal structures of both the Tet(Poly) and Hex(Poly) FeSe were also evaluated by XRD.
The Rietveld analysis against the XRD profiles reveals that the Tet(Poly) FeSe sample contains  the hexagonal-phase FeSe of 10\% as an impurity and the Hex(Poly) FeSe sample has a single phase with in the limits of the resolution. 
The Tet(Single) FeSe and Tet(Poly)  FeSe have the transition temperature  $T_c^{\rm zero}$ $\simeq$6~K estimated by  $\rho - T$ and $T_c\simeq$8~K estimated by $\chi - T$ measurements under ambient pressure, respectively, which are similar to reported values.~\cite{Hsu_08,McQueen_09,Mizuguchi_08}
This implies the present samples have the  selenium defect of  few percent.

SXPES  was carried out at both the Figure-8 undulator SX beamline BL27SU in SPring-8 using the SPECS PHOIBOS 150 hemispherical electron energy analyzer and  the twin-helical undulator 
SX beamline BL25SU in SPring-8 using  GAMMADATA-SCIENTA SES-200 spectrometer.~\cite{Ohashi_01,RSI25}
The highest total energy resolution $\Delta E$ [the full width at half maximum (FWHM)] was set to 75~meV at h$\nu$=600~eV.
HAXPES was performed at the beamline BL19LXU in SPring-8 with MB Scientific A1-HE spectrometer.
The linearly polarized light was delivered from an in-vacuum 27 m long undulator.~\cite{Yabashi_01}
The $\Delta E$ for the valence-band PES spectrum near $E_F$ was set to $\sim$380~meV at h$\nu\simeq$8~keV.
Clean surfaces of the samples were obtained by fracturing samples {\it in situ} in UHV ($\sim$4$\times10^{-8}$ Pa) at the measuring temperature ($T\simeq$20~K).

\section{results and discussion}


\begin{figure}
\includegraphics[width=7.5cm,clip]{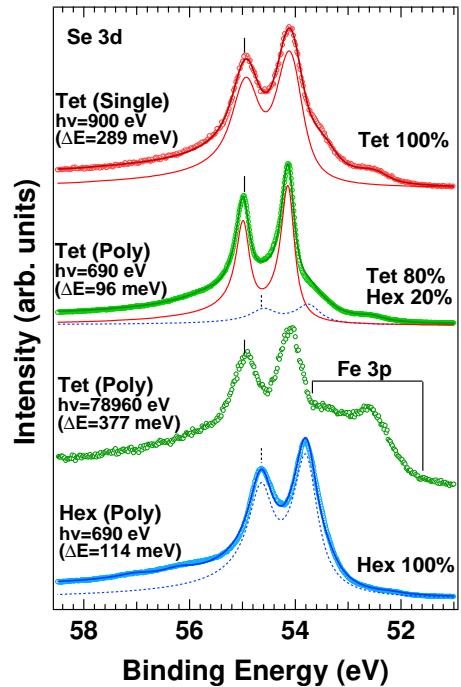}
\caption{
(Color online) Se $3d$ core-level PES spectra of FeSe.
Circles and thick solid curve indicate the experimental and fitted spectra, respectively.
The Se $3d$ components of tetragonal and hexagonal phases are shown by thin solid and broken curves.
The other components, for instance, Fe $3p$ and background are not displayed for simplicity.
}
\label{Fig_1}
\end{figure}

\begin{figure}
\includegraphics[width=8.0cm,clip]{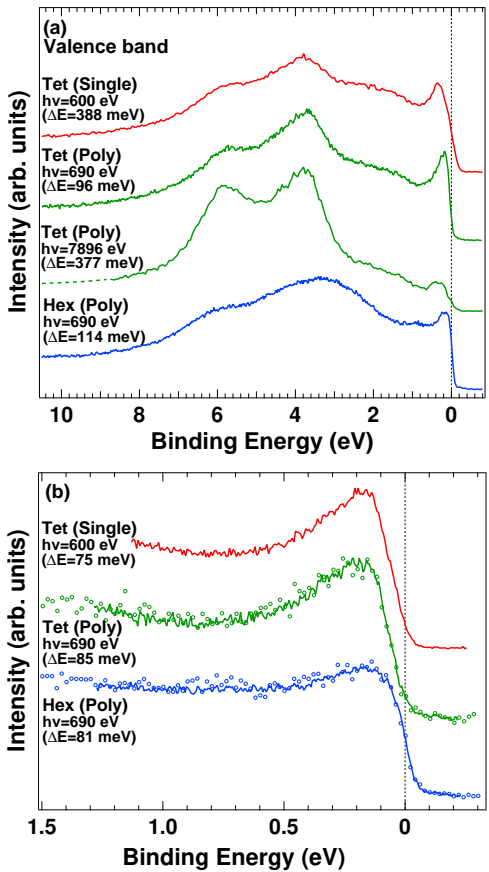}
\caption{
(Color online) Valence-band PES spectra of FeSe.
(a) Overall valence-band SXPES and HAXPES spectra.
These spectra are normalized by the area under the curves after subtracting the Shirley-type background.~\cite{Shirley_72}
(The dotted line indicates the extrapolated one assuming the Lorentzian line shape in order to normalize the HAXPES spectrum.)
(b) High-resolution PES spectra near $E_F$. 
The spectra of polycrystalline samples are normalized so that  the spectral intensity  agrees with the intensity in each normalized high-resolution overall valence-band PES spectrum (shown partly by dots).
}
\label{Fig_2}
\end{figure}

\begin{figure}
\includegraphics[width=7.5cm,clip]{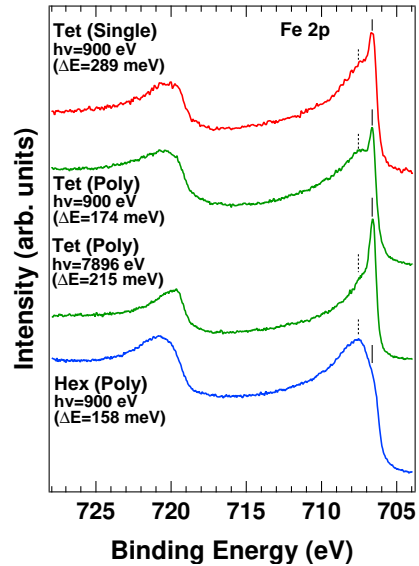}
\caption{
(Color online) Fe $2p$ core-level SXPES and HAXPES spectra of  FeSe.
}
\label{Fig_3}
\end{figure}

\begin{figure}
\includegraphics[width=7.5cm,clip]{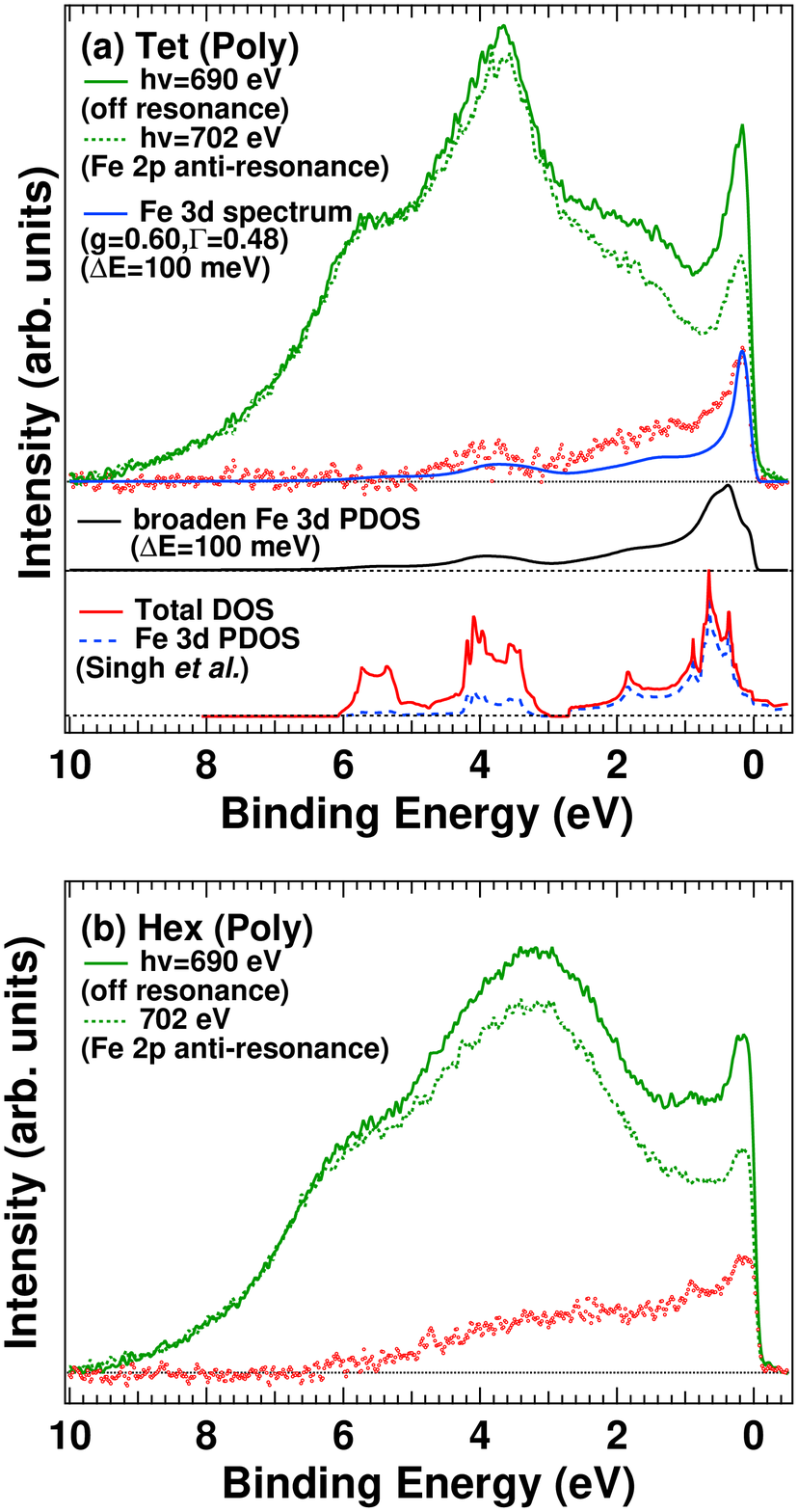}
\caption{
(Color online) Valence-band PES spectra of FeSe measured at two different photon energies.
The spectrum indicated by dots is difference between the off-resonance (measured at h$\nu$=690~eV) and anti-resonance  (measured at h$\nu$=702~eV) spectra for (a) tetragonal FeSe and (b) hexagonal FeSe.
In the middle of Figure (a), calculated Fe $3d$ spectrum including the self energy correction and the broadened Fe $3d$ PDOS are shown by solid lines.
The original DOSs calculated by Singh {\it et al.} are also shown.~\cite{Subedi_08}
}
\label{Fig_4}
\end{figure}

\begin{figure}
\includegraphics[width=8cm,clip]{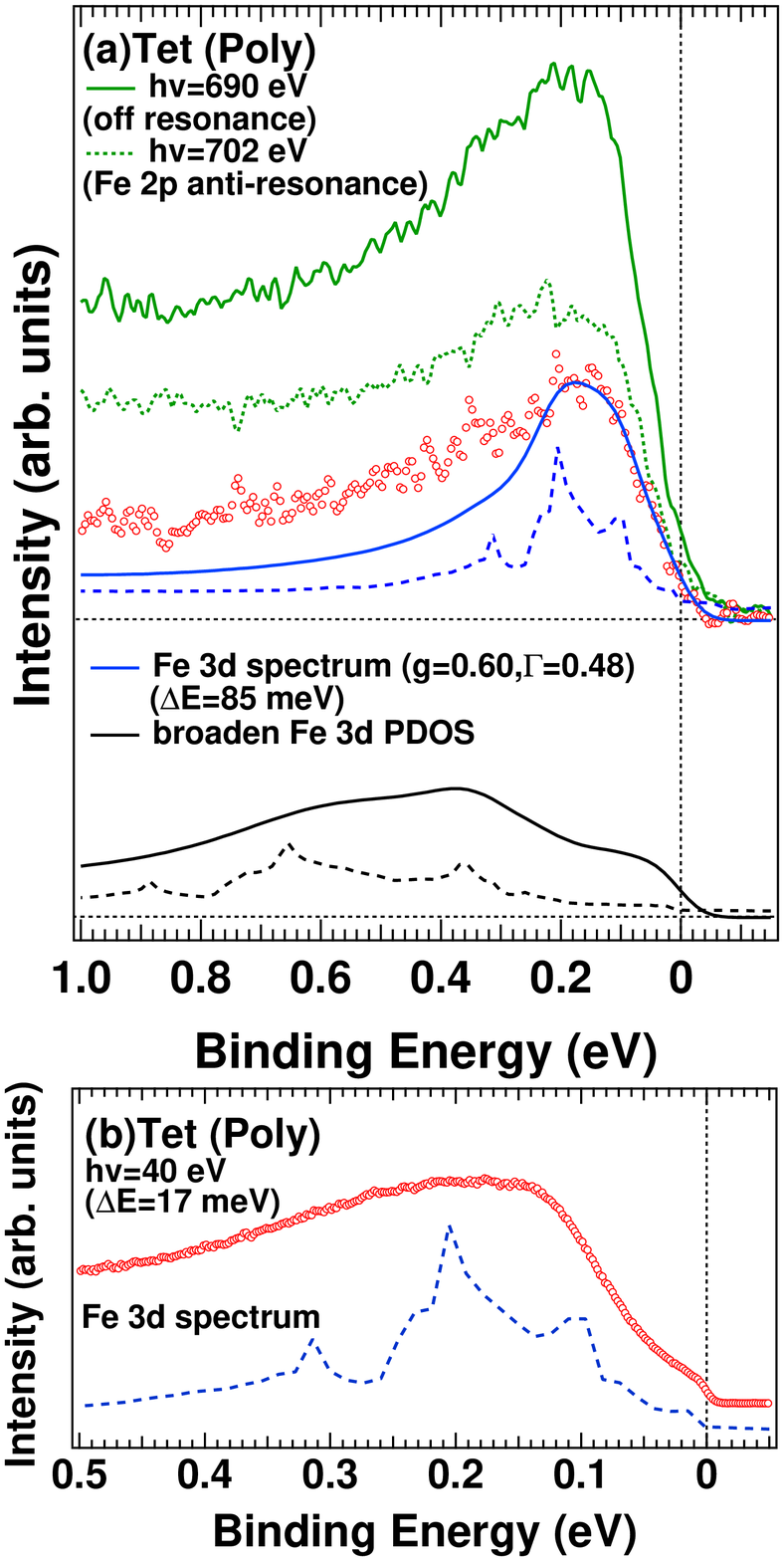}
\caption{
(Color online) High-resolution PES spectra near $E_F$. 
(a) The broken lines indicate the calculated Fe $3d$ spectrum (blue) and Fe $3d$ PDOS (black) without broadening and finite temperature effects.
Other explanations are the same as in Fig.\ref{Fig_4} (a).
(b) The PES spectrum measured at h$\nu$=40~eV.
The calculated Fe$3d$ spectrum without broadening is also shown.
}
\label{Fig_5}
\end{figure}

FeSe has two stable crystal phases, that is, the tetragonal and the hexagonal phases and often contains both in one sample as has been mentioned above.
In order to investigate the crystal phase mixing  in samples, 
Se $3d$ (including Fe  $3p$) core-level PES spectra were measured for Tet(Single), Tet(Poly), and Hex(Poly) FeSe samples. 
The Se $3d$ core-level PES spectra in both crystal phases excited by SX have a sharp doublet peak structure originating from Se $3d_{5/2}$ and $3d_{3/2}$ components as shown in Fig.~\ref{Fig_1}.
There is no surface-derived peak and  only weak plasmon satellites are seen on the higher binding energy ($E_B$) side of the Se $3d$ peaks, indicating that the surface contributions to the spectra can be ruled out in the SXPES for the shallow core levels (including the valence band) for FeSe.
In addition, the spectra in the tetragonal FeSe have the shoulder structure on the lower  $E_B$ side, which originates from the Fe $3p$ states.
In the HAXPES  the intensity of the shoulder is enhanced since
the photoionization cross section $\sigma$ of Fe $3p$ states relative to that of Se $3d$ states increases by a factor of 8 compared with the SXPES.~\cite{Lindau}
We have obtained the tetragonal and/or hexagonal components in each sample by a deconvolution procedure.~\cite{Sekiyama04}
The Se $3d$ peaks in the tetragonal and hexagonal components are located at  the certain energies in these samples, respectively, as shown by solid and dotted bars in Fig.~\ref{Fig_1}.
The peak  of the tetragonal  component is located at the $E_B$ which is about 300~meV  higher than that of the hexagonal one.
This shift is caused by  the structural difference, in other words, the difference of the covalent bond strength between these compounds, being consistent with what Se $4p$ states have a different structure in the valence band of these compounds (discussed later). 
We note that the Tet(Poly) FeSe has the hexagonal component of 20~\% in the Se $3d$ core-level PES spectrum.
The influence of the hexagonal FeSe inclusion on other spectra of Tet(Poly) FeSe will be discussed later.

%
%
%
%
%


Figure \ref{Fig_2} (a) shows valence-band SXPES and HAXPES spectra 
of  the  tetragonal and hexagonal FeSe.
In both  the tetragonal and hexagonal  FeSe  the SXPES spectra have a sharp peak in the vicinity of $E_F$.
In addition, there are some broad peaks and hump structures in the higher $E_B$ region.
The spectra of the tetragonal FeSe have the features very similar to the  recently reported result.~\cite{Yoshida_09}
These structures consist of Fe $3d$ and Se $4p$ states as has been revealed later by comparison with theoretical spectra.
In the HAXPES spectrum, it is found that the peak near $E_F$ is suppressed and the structures in the $E_B$=3-7~eV become dominant.
Such a variation of the spectral shape in the SXPES and HAXPES does not mainly originate from the increase of the bulk sensitivity but from the ratio of $\sigma$ between Fe $3d$ and Se $4p$ states,
that is, Fe($3d$)/Se($4p$) is 2.8 at h$\nu\simeq$690~eV and 0.033 at $\simeq$8~keV.
This indicates that the HAXPES spectrum of the valence band is nearly equivalent to the Se $4p$ (including weak and broad $s$) states.

Now we focus on the similarity and difference of the electronic structures between the tetragonal and hexagonal FeSe.
They have rather different Se $4p$ electronic structures between $E_B$= 3~eV and 5~eV as recognized in Fig.~\ref{Fig_2} (a). 
Furthermore,  high-resolution PES spectra reveal that Fe $3d$ states near $E_F$ also have  different features as shown in Fig.~\ref{Fig_2} (b).
We note that each spectrum of Tet(Poly) and Hex(Poly) FeSe is normalized by the integrated intensity of its high-resolution whole valence-band PES spectrum after subtracting the background.
The Tet(Poly) FeSe has a prominent peak at $E_B\simeq$ 180~meV.
The photoemission intensity decreases steeply toward $E_F$ and is very weak at $E_F$.
Meanwhile, the Hex(Poly) FeSe has a less prominent peak, which is closer to $E_F$.
In addition, the obvious Fermi cut off is observed in the Hex(Poly) FeSe.
Here we would stress that the spectra near $E_F$ in the Tet(Single) FeSe is in good agreement with in the Tet(Poly) FeSe,
indicating that the influence of the hexagonal FeSe inclusion in Tet(Poly) FeSe is negligible near $E_F$.


Fe $2p$ core-level PES spectra of the tetragonal and hexagonal FeSe are shown in Fig.~\ref{Fig_3}.
There is no charge-transfer satellite, as has been pointed out  in  LaFeAsO.~\cite{Malaeb_08}
In addition, one can see the sharp-peak-and-shoulder structure in the $2p_{3/2}$ component of the SXPES
spectra of both the Tet(Single) and Tet(Poly) FeSe as indicated by solid and dotted bars.
Considering that  the sharp peak becomes more prominent in the HAXPES spectrum,
 the peak can be assigned to the bulk $|2p^53d^6\rangle$ component.
 The sharp peak structure  suggests the existence of the highly coherent valence electrons in the bulk.
Then, the shoulder structure is attributed to 
the photoemission from the surface (and grain boundaries in the polycrystalline sample).
These indicate that the photoelectrons with the low $E_K$ (of  less than a few hundred eV) lead to the large surface contribution to the spectrum even in the SXPES experiment.~\cite{Tanuma_94} 
The possibility  cannot be ruled out that the slightly stronger shoulder intensity in the SXPES spectrum of Tet(Poly) FeSe than that of Tet(Single) FeSe originates from the influence of the hexagonal FeSe inclusion.
Meanwhile, the Hex(Poly) FeSe has a broad peak originating from multiplet structures.
The overall spectral shape is very similar  to the reported spectrum in the hexagonal Fe$_7$Se$_8$.~\cite{Shimada_98}


Let us discuss  Fe $3d$ states in the valence band PES spectra of the  tetragonal and  hexagonal FeSe.
In Fig.~\ref{Fig_4} (a), the valence-band PES spectra of the Tet(Poly) FeSe measured at two different photon energies are shown.
The solid line indicates the spectrum obtained at h$\nu$=690~eV, which is labeled as  ``off-resonance'' in comparison to another spectrum.
The dotted line shows the spectrum which was measured at the energy just below the threshold of Fe $2p$-$3d$ absorption edge, labeled as ``Fe $2p$ anti-resonance''.
In this spectrum Fe $3d$ states are strongly suppressed since the tuned photon energy corresponds to the energy providing  the mostly minimum transition probability in the Fano lineshape.~\cite{Fano}
One can see  that the spectral weight between $E_F$ and $E_B$=3~eV is remarkably reduced.
This indicates  Fe $3d$ dominant states are located in this $E_B$ range.
However, some spectral weights still remain, suggesting that there are Se $4p$ states hybridized with the Fe $3d$ states  as seen in the HAXPES spectrum of Fig.~\ref{Fig_2} (a).
Meanwhile,  no significant reduction above $E_B$=3~eV is seen except for the slight decrease around $E_B$=4~eV, since these structures mainly originate from Se $4p$ states.
We note that one often employs the (on-) resonant PES  to investigate the contribution of the specific electronic states by using the photon energy tuned to the core-level absorption maximum.
In the present case, however, the Auger decay process becomes dominant and the valence-band structures are smeared out due to the large background.
The difference spectrum between the off-resonance (h$\nu$=690~eV) and anti-resonance spectra is also shown as dots in Fig.~\ref{Fig_4} (a), which represents the Fe $3d$ partial density of states (PDOS) if the electron correlation effect  is negligible.
It has a sharp peak in the vicinity of $E_F$ and broad hump structures at around $E_B$=1.5~eV and 4~eV.
The difference spectrum has similar features to the experimentally obtained Fe $3d$ states of LaFe{\it Pn}O$_{0.94}$F$_{0.06}$ ({\it Pn}=P, As).~\cite{Malaeb_08}
Meanwhile, the Hex(Poly) FeSe has much different Fe $3d$ states  from the Tet(Poly) FeSe as shown in Fig.~\ref{Fig_4} (b).
The difference spectrum has a large spectral weight up to $E_B$=6~eV and broad hump structure around 3~eV.
These features are consistent with those in Fe$_7$Se$_8$ spectrum measured at h$\nu$=100~eV.~\cite{Shimada_98}
The SX spectrum has a stronger peak near $E_F$ than the lower-h$\nu$ spectrum presumably due to the high bulk sensitivity.~\cite{Sekiyama_nature}

The results of the band structure calculations with the local-density approximation (LDA) for the tetragonal FeSe done by Singh {\it et al.} and the Fe $3d$ PDOS broadened by Gaussian and
Lorentzian functions representing the experimental energy resolution and the lifetime broadening effect are also shown in the  Fig.~\ref{Fig_4} (a).
By comparison with them,  narrowing the Fe $3d$ band width and the energy shift of the band toward $E_F$ are found in the difference spectrum of the Tet(Poly) FeSe, suggesting that the electron correlation effect cannot be negligible to discuss the electronic structures in this system.
In order to take into account  the electron correlation effect quantitatively, the self energy ${\mathit\Sigma}({\boldsymbol k}, \omega)$ defined as ${\mathit\Sigma}({\boldsymbol k}, \omega)\equiv G_0^{-1}({\boldsymbol k}, \omega)-G^{-1}({\boldsymbol k}, \omega)$ is often employed, where $G_0({\boldsymbol k}, \omega)$ and $G({\boldsymbol k}, \omega)$ are the  one-electron Green's functions without and with the electron-electron interaction.
We have calculated the  spectrum with the self energy correction in accordance with the procedure reported by Shimada {\it et al.}~\cite{Shimada_98}
The ${\boldsymbol k}$-independent self energy is assumed to be given by ${\mathit\Sigma}(\omega)=g\omega/(\omega+i{\mathit \Gamma})^2$, where
$g$ and ${\mathit \Gamma}$ are employed as  fitting parameters and $\hbar\omega<\mu(=0)$ for the occupied state.
The ${\boldsymbol k}$-integrated one-particle spectral function for Fe $3d$ states including the electron-electron interaction, hereafter called ``Fe $3d$ spectrum'' $A_d(\omega)$, is obtained as follows,
\begin{eqnarray}
A_d(\omega)\!\!\!&=&\!\!\!\sum_{\boldsymbol k}A_d({\boldsymbol k},\omega)  \\
\!\!\!&=&\!\!\!-\frac{1}{\pi}\sum_{\boldsymbol k} \int_{-\infty}^{+\infty} d\epsilon_{\boldsymbol k}^{0}D_d(\epsilon_{\boldsymbol k}^0){\rm Im}G({\boldsymbol k}, \omega) \nonumber \\
\nonumber \\
&=&\!\!\!-\frac{1}{\pi}\sum_{\boldsymbol k}\int_{-\infty}^{+\infty} d\epsilon_{\boldsymbol k}^{0}\biggl[ D_d(\epsilon_{\boldsymbol k}^0) \nonumber\\
\!\!\!&\times&\!\!\! \frac{{\rm Im}{\mathit \Sigma}({\boldsymbol k}, \omega)}{\{\hbar\omega-\epsilon_{\boldsymbol k}^0-{\rm Re}{\mathit \Sigma}({\boldsymbol k}, \omega)\}^2+\{{\rm Im}{\mathit\Sigma}({\boldsymbol k}, \omega)\}^2}\biggr], \ \ \ \ \ \ \  \nonumber
\end{eqnarray}
where  $D_d(\epsilon_{\boldsymbol k}^0)$ denotes Fe $3d$  PDOS.
The obtained spectrum after optimizing the parameters ($g$=0.60, ${\mathit\Gamma}$=0.48) is shown in Fig.~\ref{Fig_4} (a), which includes the broadening and finite temperature effects.
The calculated Fe $3d$ spectrum has the peak near $E_F$ as a coherent quasi-particle peak, which becomes much sharper than in the PDOS without the electron correlation effect, and has two broad hump structures at around 2 and 4~eV.
Thus, it reproduces well the overall valence-band features experimentally obtained.
The comparison between the experimental and calculated Fe $3d$ states near $E_F$ are shown in Fig.~\ref{Fig_5}.
The peak shift of the spectrum toward $E_F$ due to the self-energy correction is obviously seen, the energy of which corresponds to ${\rm Re}{\mathit \Sigma}(\omega)$ in the non-broadened spectrum.
One can also see that the leading edge behavior in the vicinity of $E_F$  and the peak position is well reproduced.
We have measured the further higher-resolutioin spectrum at a low energy excitation (h$\nu$=40~eV) shown in Fig.~\ref{Fig_5} (b).
In this photon energy the Fe $3d$-state-dominant spectrum is obtained since the $\sigma$ of Fe $3d$ states is 16 times larger than that of Se $4p$ states.~\cite{Lindau}
Although the spectrum should be very sensitive to the surface electronic structures, the spectral features agree well with the calculated Fe $3d$ spectrum.

\begin{table*}
\caption{
Experimentally and theoretically obtained mass enhancement factor  ($Z^{-1}$ or $m^*/m_b$)  for typical Fe-based superconductor systems.
$\dagger$ indicates the non-superconductor.
The Sommerfeld coefficient $\gamma$ is also listed.
}
\begin{ruledtabular}
\begin{center}
\begin{tabular}{lccccc}
 {} &LaFe{\it Pn}O & $^\dagger$BaFe$_2$As$_2$ & (Ba,K)Fe$_2$As$_2$ & FeSe & Fe(Se,Te)\\ \hline
$m^*/m_b$\ [PES]&{1.5-2.2 ({\it Pn}=P), 1.8 ({\it Pn}=As, F-doped)} &{1.5} &{2.7}&{3.6} &{6-20}\\
{}&Ref.[{\onlinecite{Lu08,Qazilbash09, Malaeb_08}}]&{Ref.[{\onlinecite{Yi09}}]}&{Ref.[{\onlinecite{Yi09}}]}&{}&{Ref.[{\onlinecite{Tamai10}}]}\\
\\
$m^*/m_b$\ [DMFT]&{1.9-2.2({\it Pn}=P), 1.6 ({\it Pn}=$^\dagger$As)} &{1.8-2.1} &{--}&{2-5} &{--}\\
{}&{Ref.[{\onlinecite{Skornyakov10,Aichhorn09}}]}&Ref.[{\onlinecite{Skornyakov09}}]&{}&Ref.[{\onlinecite{Aichhorn10}}]&{}\\
\\
$\gamma$\ (mJ/mol K$^2$) &{10.1-12.5({\it Pn}=P)} &{16-37} &{23} &{5.4-9.1} &{39}\\
{}&Ref.[{\onlinecite{Suzuki09, McQueen08}}]&Ref.[{\onlinecite{Rotter08, Ni08}}]&Ref.[{\onlinecite{Ni08}}] &Ref.[{\onlinecite{Hsu_08, McQueen_09}}]&{Ref.[{\onlinecite{Sales09}}]}\\

\end{tabular}
\end{center}
\end{ruledtabular}
\label{Table_summary}
\end{table*}


There is a discrepancy between the experimentally and theoretically obtained Fe $3d$ spectra in the $E_B$=0.2-3~eV as seen in Fig.~\ref{Fig_4}(a) and Fig~\ref{Fig_5}(a).
This should be due to the contribution of the incoherent part of the spectral function, $A_d^{\rm inc}({\boldsymbol k}, \omega)$.
In the present analysis, only the pole part of the Green's function, that is, the coherent part is considered.
Thus, the quasi-particle spectral weight is reduced to $Z_{\boldsymbol k}$, 
\begin{equation}
\int_{-\infty}^{+\infty} d\omega A_d^{\rm coh}({\boldsymbol k},\omega)=Z_{\boldsymbol k}(<1)
\end{equation}
where $Z_{\boldsymbol k}$ and $A_d^{\rm coh}({\boldsymbol k},\omega)$ are the renormalization factor and the  coherent part of the spectral function $A_d$ $({\boldsymbol k}, \omega)$ for Fe $3d$ states.
According to the sum rules of the spectral function, the following spectral weight of the incoherent part remains in the occupied and unoccupied states,
\begin{equation}
\int_{-\infty}^{+\infty} d\omega A_d^{\rm inc}({\boldsymbol k},\omega)=1- Z_{\boldsymbol k}.
\end{equation}
The incoherent part should appear  at  $\hbar\omega\simeq\epsilon_{\boldsymbol k}^0+{\rm Re}{\mathit \Sigma}(\omega)$ on the higher $E_B$ side of  the quasi-particle peak, for
instance, $\hbar\omega\simeq1.0$eV for $\epsilon_{\boldsymbol k}^0\simeq$0.65~eV in the largest peak of  the original Fe $3d$ PDOS.
This is consistent  with the energy region in which the disagreement between experimental and calculated spectra are seen.
In fact, in very recent articles, theoretical works predict that the incoherent spectral weight appears around $E_B\simeq$2~eV as a lower Hubbard band.~\cite{Miyake10,Aichhorn10}

The $Z_{\boldsymbol k}^{-1}$ at $k=k_F$, that is, $Z^{-1}$ depends on the real part of the self energy ${\rm Re}{\mathit\Sigma}(\omega)$ as follows,~\cite{Imada_98}
\begin{equation}
Z^{-1}\equiv1-\frac{\partial {\rm Re}{\mathit\Sigma}(\omega)}{ \partial \omega}\biggr|_{\omega=0}=1+\frac{g}{\mathit \Gamma^2}.\\
\end{equation}
The $Z^{-1}$ represents the mass enhancement due to the band narrowing of  the quasi-particle peak.
In the present work, we have obtained  $Z^{-1}$=3.6, which is about  twice as heavy as the reported values in LaFeAsO$_{0.94}$F$_{0.06}$.~\cite{Malaeb_08}
In Table~\ref{Table_summary},  the obtained $Z^{-1}$, that is, $m^*/m_b$ (the ratio of the enhanced effective mass to the bare band mass) and some reported values related to the electron correlation in Fe-based superconductor systems are summarized.
Spectroscopic and theoretical studies clearly indicates the common tendency of the electron correlation strength in these systems, that is, stronger on the right side, although the $\gamma$ value in FeSe is peculiarly smaller than in other superconductors.
We conclude that the electron correlation in the FeSe superconductor is relatively strong, being quantitatively consistent with the results of the DFT +DMFT (the density-functional theory combined with the dynamical mean-field theory) study.~\cite{Aichhorn10}

\section{summary}

We have performed the bulk-sensitive SXPES  and HAXPES for Fe-based superconductor FeSe.
It is suggested in the Fe $2p$ core-level PES that the tetragonal FeSe has the Fe $3d$ electrons with the highly coherent  character unlike the non-superconducting FeSe with the hexagonal structure. 
In comparison to the results of band structure calculations,  the Fe $3d$ band narrowing and its energy shift  are revealed, which originates from the electron correlation effect.
Considering the self energy correction, the larger mass enhancement  ($Z^{-1}\simeq$3.6) than in other Fe-As superconductors are obtained.
In addition,  the incoherent part of the quasi-particle spectrum are found and successfully separated.  
The obtained mass enhancement value and the energy position of the incoherent part are consistent with the results of recent dynamical mean-field calculations.


\ \
\begin{acknowledgments} 

We would like to thank K. Sugimoto, K.~Mima, R.~Yamaguchi, Y.~Miyata  for supporting experiments and 
D. J. Singh for providing the results of band structure calculations.
The  experiments  were performed at SPring-8 with the approval of the Japan Synchrotron Radiation Research Institute (JASRI)
(Proposal No.~2008B1149 and No.~2009A1122)
under the support of Grant-in-Aid for 
``Open Research Center'' Project
from the Ministry of Education, Culture, Sports, Science, and Technology, Japan, and 
Research Foundation for Opto-Science and Technology,
the Sumitomo Foundation, Research Institute of Konan University, and 
the Hirao Taro Foundation of the Konan University Association.

\end{acknowledgments}



\end{document}